\documentclass[twocolumn,showpacs,preprintnumbers,amsmath,amssymb]{revtex4}

\usepackage{graphicx}
\usepackage{dcolumn}
\usepackage{bm}
\usepackage{amssymb}

\newcommand\ztg{Z \to \gamma \gamma}

\newcommand\zptg{Z' \to \gamma \gamma }

\newcommand\zp{Z'}
\newcommand\tg{\gamma \gamma}

\newcommand\zptmu{Z'\to \mu^+ \mu^-}

\def\address{\@ifstar{\address@star}%
  {\@ifnextchar[{\address@optarg}{\address@noptarg}}}

\begin{document}

\author{S.N.~Gninenko$^a$}
\author{A.Yu.Ignatiev$^b$}
\author{V.A.~Matveev$^a$}

\affiliation{$^a$Institute for Nuclear Research of the Russian Academy of Sciences, Moscow 117312\\
$^b$Theoretical Physics Research Institute, Melbourne 3163, Australia}


\title{  Two photon decay of $Z'$ as a probe of Bose symmetry violation  at the CERN LHC}

\date{\today}

\begin{abstract} 
The question if the Bose statistics is broken at the TeV scale is discussed. 
The  decay  of a new heavy spin 1 
gauge boson $Z'$ into two photons,  $\zptg$, is forbidden by the Bose statistics
among other general principles of quantum field theory (Landau-Yang theorem). We point out that the search for this decay can be effectively used to probe the Bose symmetry violation at the CERN LHC.
\end{abstract}

\maketitle


\section{Introduction}

For all known particles, there is a remarkable one-to-one correspondence between their spin type and statistics type:
\begin{eqnarray}
\nonumber
\rm integer \;spin \leftrightarrow commutation\; relations \leftrightarrow \\  \rm exchange- symmetric \; wave \; functions \leftrightarrow \\  \nonumber \rm Bose\; statistics;
\end{eqnarray}
\begin{eqnarray}
\nonumber
\rm half-integer \;spin \leftrightarrow anticommutation\; relations \leftrightarrow \\  \rm exchange-antisymmetric \; wave \; functions \leftrightarrow \\ \nonumber \rm Fermi\; statistics.
\end{eqnarray}

In quantum field theory, this spin-statistics 
connection (SSC)
can be understood in different ways.
For free fields, the connection  was established   using conventional lagrangian and group-theoretical methods \cite{w}.   

Interacting fields were tackled  using axiomatic quantum field theory  (without any lagrangians), and by early 1960s the celebrated spin-statistics theorem was proved \cite{sw}. 

It is worth recalling that the axiomatic treatment of the spin-statistics connection \cite{sw}  {\em does not} cover the case of quantum electrodynamics and other gauge theories including the Standard Model (for a detailed explanation, see \cite{b}, Sec. 8.1).  

The physical root of this difficulty is the fact that while the photon is a massless spin-1 particle with only {\em two} polarizations,  the corresponding 4-potential $A_{\mu}$ has  {\em four} components.

Consequently, ordinary Hilbert space with {\em positive} metric cannot accomodate the 4-component field operator $A_{\mu}$, and the introduction of {\em indefinite-metric} space is required. 

This  is at odds with  the axiom underlying the proof of the  spin-statistics theorem,
which states that the Hilbert space metric must be {\em positive}.

As a result, the original set of axioms  \cite{sw} needs to be modified to include   Hilbert spaces with {\em indefinite} metric as well. Although progress has been achieved in this area (see, e.g., \cite{b}, Ch.10)
the extension of the spin-statistics theorem  to gauge theories is yet to be formulated.


In any case, the theorem does not forbid small
violation of the SSC and  the question whether it exists or not 
remains open. (That does not mean, however, that construction of a theory with small statistics violation is easy. To date, most attempts to find  a local relativistic quantum field theory with small statistics violation have been unsuccessful.)

There have been several works that have attempted to improve the original spin-statistics theorem by going beyond the Bose-Fermi alternative \cite{other}, and, in particular, to rule out the small violation of SSC. However, they also assume positivity of Hilbert space metric and hence do not apply to gauge theories. 

Even if the small SSC violation was shown to be theoretically forbidden on the basis of general principles of QFT such as Lorentz invariance, locality etc., the experimental tests of SSC would still be of interest  because they would be important tests of   those general principles.

We may assume that at some extremal or  small enough distances where the usual notions of the local fields become invalid, their  proper reconsideration and a generalisation should be required,  which can bring us  the possibility of changing or correcting the usual Spin-Statistics Correspondence.

Since 1987 there have been significant theoretical and experimental efforts 
to motivate and find 
tiny departure from the 
established connection between spin and statistics.

Originally, most efforts [5-30], especially in the experimental field, were actually concentrated on discussing small violation of the Pauli exclusion principle rather than violation of Bose statistics. Many dedicated experiments have been performed to give strong bounds on the violation of the Pauli principle.  Also, the topic was discussed in the context of string theories \cite{j} and cosmology \cite{s}. For recent reviews, see \cite{i}.

Later, Bose statistics came under scrutiny as well. 
Initially, experimental proposals of searching for deviations from Bose symmetry   used  the spin-zero nucleus of oxygen $^{16}O$ as the test object \cite{h, t}.

For a review of subsequent experiments with $O_2$ and $CO_2$ molecules, see \cite{t1}.

The first experimental upper limit on the validity of Bose statistics for photons was obtained in Ref. \cite{ign} based on the idea of forbidden two-photon decays of Z-boson.

The same idea was later explored in Ref. \cite{dbd},  in the context of low-energy, high-sensitivity atomic two-photon transitions   (see also \cite{bm}).


 This approach exploits one  of the important  consequences of the Bose symmetry, first observed by  Landau \cite{dau} and Yang \cite{yang}:   a pair of photons cannot form  a state with total angular momentum equal to unity. The Landau-Yang theorem uses general principles of rotation invariance, gauge invariance, and Bose statistics 
 to derive certain selection rules for
decays of a parent particle into two 
photons. 
For a parent with spin one, the decay amplitude into the 
exchange-symmetric state of two photons vanishes.

Therefore, the decay of any spin-1 boson into two photons is absolutely forbidden (for a textbook proof, see, e.g., \cite{nish}).
If, however photons do not obey Bose statistics, 
there will be a nonzero decay amplitude
involving two photons  in an 
exchange-antisymmetric state \cite{ign}. This provides  a clear reason why the diphoton  system  is especially interesting in testing  the degree with which Bose symmetry is exact.  

Recent experiments have explored the possibility of 
small violations of the usual relationship between spin and 
statistics, which are  impossible 
within conventional quantum field theory.
 Experiments  searching for the transitions between atomic 
states with $J= 0$ and $J'= 1$ for degenerate photons (i.e. photons of equal energies)
test the Bose statistics at the eV scale  and yield 
upper limit on the ratio $\nu$ of the rate of  statistics-violating transitions to an equivalent statistics-allowed 
transition rate, $\nu < 4.0\times 10^{-11}$ at the 90\% confidence level \cite{bud}.
 At higher energies  the limits on the branching fraction of two-photon decays of the triplet positronium (orthopositronium) $Br(oPs\to \tg)\lesssim 2.4\times 10^{-5}$ \cite{ops} and 
charmonium  $Br(\chi_{c1}\to \tg)\lesssim 3.5\times 10^{-5}$ \cite{cleo}  have been reported. At the energy scale of 100 GeV 
the results of the search for the gauge boson $Z$ into two photons were obtained at LEP
\cite{leptg} $Br(\ztg) < 1.4 \times  10^{-4}$.


From  general arguments, one can expect that any violations of Bose symmetry, if any, would be better manifested at higher energy scales. 
Then the question arises: could
 interesting physics 
be found by combining  the idea of searching for Bose statistics violation and assumption
that Bose symmetry might be broken at a high energy frontier at the CERN LHC?  Out of 
all (neutral) spin 1 bosons it is natural to concentrate on the heaviest one- the new
heavy gauge boson $Z'$ which appears naturally in many extensions of the SM.

The explicit goal of this paper is to extend the results of the previos work [37] to the TeV scale and to show  
 that the search for the decay $Z'\to \gamma \gamma$ at the CERN LHC
could  result in  a more radical departure from standard physics: the possible observation of a small violation of Bose statistics, which  would provide  strong evidence for the existence of new physics.
The rest of the paper is organized as follows:
in Sect.II we describe the phenomenological model \cite{ign} of the Bose symmetry violation  and its extensions; we then
show how the parameters of the model 
 are constrained by theoretical arguments and electroweak precision data. Following this, Sec. III deals with the $Z'$ sector in some detail.  The main characteristics of the $Z'$ boson are described within the framework of several $Z'$ models. Section IV-VI
present the results of a sensitivity study for the Bose symmetry violating 
process $pp\to Z'\to \tg$ at the CERN LHC
at $\sqrt{s}=14$ TeV.    Finally, Sec. VII presents the conclusions of this work.

\section{A model for Bose symmetry violation}

In this section we describe a simple model of Bose symmetry violation suggested in Ref. \cite{ign} and its extensions.


The method \cite{ign} is to write down the most general form of 
the decay amplitude of the spin-1 particle into two photons and then apply the conditions of 
gauge invariance and Bose symmetry to that amplitude. If both conditions are applied, the 
resulting amplitude is exactly zero. However, if we impose the condition of 
gauge invariance but do not require the Bose symmetry, the resulting amplitude is not zero. 
 We then can obtain the two-gamma decay rate of Z-boson and 
compare it to the experimentally known upper bound on the branching ratio of $\ztg$. 
In this way  a direct bound on Bose symmetry 
violation for photons can be obtained \cite{ign}.
 
Now, a few remarks about the relation of this method to alternative models of small 
Bose symmetry violation. The most well-studied model is "the quon model" proposed in   \cite{green1}.
Quons 
are particles described by the commutation relations of the form: 
\begin{equation}
a_k a_l^+ - qa_l^+l a_k = \delta_{kl}.
\label{com} 
\end{equation}
However, it was shown in Ref.\cite{cg} that in relativistic field theories quons must be either fermions or bosons. For this reason, the quon theory cannot be used for the analysis of the decay $\ztg$. 



Let us turn now to  the construction of $\ztg$ decay amplitude. 
We require that this amplitude satisfies all the standard conditions, such as relativistic 
invariance and gauge invariance, but we do not require this amplitude to be symmetric 
under the exchange of photon ends. 
The most general Lorentz invariant form of the amplitude $S$ of the decay $\ztg$ is: 
\begin{equation}
S(k_1,k_2,\epsilon_1, \epsilon_2)=c_{\lambda \mu \nu} (k_1,k_2)\epsilon_0^\lambda \epsilon_1^\mu \epsilon_2^\nu 
\label{ampl}
\end{equation}

where $k_1$ and $k_2$ are photon momenta, $\epsilon_1$ and $\epsilon_2$ are photon polarization vectors, $\epsilon_0$ is $Z$-boson 
polarization vector. 

Even though violation of SSC could require violation of Lorentz invariance as well, using more general parametrization in Eq. (\ref{ampl}) would be overkill.

Note that terms in $c_{\lambda \mu \nu}$ proportional to $k_{1\mu}, k_{2\nu}$ and $k_{1\lambda} + k_{2\lambda}$ do not contribute to $S$  due to  
the conditions 
\begin{equation}
\epsilon_1^\mu k_{1\mu} = 0,~\epsilon_2^\nu k_{2\nu} = 0,~ \epsilon_0^\lambda ( k_{1\lambda} + k_{2 \lambda}) = 0
\label{polar}
\end{equation}
and can 
therefore be ignored.

 We focus first on the part of $c_{\lambda \mu \nu}$ that does not contain the $\epsilon$-tensor: it has the following Lorentz invariant form: 
\begin{eqnarray}
c_{\lambda \mu \nu} = b_1 g_{\lambda \mu} k_{1\nu} \\ \nonumber
+b_2 g_{\lambda \nu} k_{2\mu} + g g_{\mu \nu }(k_{1\lambda}- k_{2\lambda})+ h(k_{1\lambda}- k_{2\lambda})k_{1\nu}  k_{2\mu} 
\label{gauge}
\end{eqnarray}

Now, the condition of the electromagnetic gauge invariance reads 
\begin{equation}
c_{\lambda \mu \nu} k_1^\mu=0, \;\; c_{\lambda \mu \nu} k_2^\nu  =0.
\label{g1}
\end{equation}


These conditions can be satisfied if we put
\begin{equation}
 b_1 = b_2 = 0,\;\;\;h=-\frac{2g}{M_Z^2}. 
\label{h}
\end{equation}

Then the amplitude $c_{\lambda \mu \nu}$ becomes \cite{ign}
\begin{equation}
c_{\lambda \mu \nu}=g(k_1 - k_2)_\lambda (g_{\mu \nu}- \frac{2k_{1\nu} k_{2\mu}}{M_Z^2}). 
\label{c}
\end{equation}

In principle, $g$ could depend on some scalar products of the momenta, but in our case, since 
all the particles are on mass shell, we have $k_1 k_2 = M_Z^2/2$ 
(and, of course, $k_1^2=k_2^2=0$), so 
that $g$ is a pure number. 
Note that the above amplitude automatically satisfies the condition 
$(k_1 + k_2 )^\lambda c_{\lambda \mu \nu} = 0$. 
We see that this amplitude, as expected, violates Bose symmetry because
\begin{equation} 
c_{\lambda \mu \nu} (k_1 , k_2 ) = - c_{\lambda \nu\mu } (k_2 , k_1 )
 \end{equation}
whereas Bose symmetry requires $c_{\lambda \mu \nu} (k_1 , k_2 ) = +c_{\lambda \nu\mu } (k_2 , k_1 )$. 

Thus the parameter $g$ can be interpreted as the parameter of Bose statistics violation which will be marked by the subscript $B1$: $g\equiv g_{B1}$. 

Next, it can be shown that the following Lorentz-invariant terms containing the $\epsilon$-tensor also satisfy the conditions of gauge invariance and Bose-antisymmetry:
\begin{equation}
\label{t1}
c_{\lambda \mu \nu} =\frac{g_{B2}}{M^2}(k_{1\lambda}-k_{2\lambda})\epsilon_{ \mu \nu \alpha\beta}k_{1}^{\alpha}k_{2}^{\beta}
\end{equation} and 
\begin{eqnarray}
\label{t2}
c_{\lambda \mu \nu} =\frac{g_{B3}}{M^2}[(\epsilon_{\lambda \mu \alpha\beta} k_{1\nu}-\epsilon_{\lambda  \nu \alpha\beta}k_{2\mu})k_{1}^{\alpha}k_{2}^{\beta}+\\\nonumber \epsilon_{\lambda \mu \nu\alpha}(k_1+k_2)^{\alpha}(k_1 k_2)].
\end{eqnarray}

Now, calculating the width of the decay $\ztg$ with the help of the amplitude Eq.(\ref{c})
we obtain \footnote{the difference between the numerical factors here and in Ref.\cite{ign} is due to a mistake in the latter.}: 
\begin{equation}
\Gamma_1(\ztg) =\frac{1}{16\pi M_Z}|S|^2=\frac{g_{B1}^2 M_Z}{24\pi}.
\label{rate1}
\end{equation}
Experimentally, it has recently been measured at LEP \cite{pdg} that 
\begin{equation}
Br(\ztg) < 5.2 \times  10^{-5}.
\label{lim} 
\end{equation}
Therefore 
\begin{equation}
\frac{\Gamma_1(\ztg)}{\Gamma_{tot}(Z)} =\frac{g_{B1}^2M_Z}{24\pi\Gamma_{tot}(Z)}<5.2 \times  10^{- 5}.
\label{rate2}
\end{equation}
Thus, taking into account that $\Gamma_{tot}(Z)\simeq 2.5$ GeV \cite{pdg}, finally, we can obtain our 
upper bound on the Bose violating coupling
\begin{equation}
g_{B1} < 1.1 \times  10^{-2}.
\label{gb1} 
\end{equation}

We see that in the framework of this  model the rate of the decay 
$Z\to \tg $ is  small; however, it can be    enhanced 
for higher-mass particles. Hence, the heavier spin-1 $Z'$  is   a good candidate 
for the searching for effect of Bose symmetry violation through the $Z'$ decay 
into two gammas. 

A similar analysis can be carried out for the amplitudes of Eq.(\ref{t1}) and (\ref{t2}). 
The decay widths due to the amplitudes (\ref{t1},\ref{t2}) are equal to: 
\begin{equation}
\Gamma_2(\ztg) =\frac{g_{B2}^2 M_Z}{96\pi},\;\;\Gamma_3(\ztg) =\frac{g_{B3}^2 M_Z}{96\pi}.
\end{equation}
The corresponding upper bounds on the constants $g_{B2}, \;g_{B3}$ are:
\begin{equation}
g_{B2}, \;g_{B3} < 2.2\times 10^{-2}.
\label{gb2}
\end{equation}

In the context of our approach, the  reason for treating the three amplitudes separately is purely technical and does not involve  the essential physics. Indeed, adding the three amplitudes with three arbitrary coupling constants would be possible but it would only add considerable complexity and obscurity into experimental simulations without any gains in physical understanding.

A few remarks are now in order concerning the relationship between the approaches developed in Ref.  \cite{ign} and a later work, Ref. \cite{dbd}.

In Ref. \cite{dbd} it was claimed that the limit obtained in \cite{ign} is too weak to be of any significance. This conclusion was reached within a very specific model for Bose symmetry violation which is different from the model-independent approach suggested in \cite{ign}. A detailed discussion of similarities and differences between the two frameworks would be out of place in the present paper, so we restrict ourselves to a few general comments only.

In a nutshell, the difference is this: Instead of constructing the most general Bose-violating amplitude of the decay and the coupling constant $g_B$, the authors of Ref.\cite{dbd} introduce the ``probability for two photons to be in an antisymmetric state'' $\nu$. The physical decay width $\Gamma$ is then obtained as $\Gamma=\Gamma_B+\nu \Gamma_F$, where $\Gamma_B$ is the width for the  decay into ordinary ``bose-photons'', and $\Gamma_F$ is the width of Bose-symmetry violating decay into ``fermi-photons''.
For instance, if the decay amplitude for the usual, ``bose-photons''  1 and 2 is $A_B=A_{12}+A_{21}$, then the corresponding amplitude for ``fermi-photons'' would be $A_F=A_{12}-A_{21}$, with  $\Gamma_B\propto |A_B|^2$ and  $\Gamma_F\propto |A_F|^2$.

This approach may look simple and natural, but five points should be kept in mind:

1. Well-known difficulties arise when $\nu$ is introduced.   According to quantum mechanics, every probability is the square of (modulus of ) amplitude. So, after we introduced $\nu$ we must introduce the ``amplitude for two photons to be in an antisymmetric state'', let us call it $\mu$. 

So, the Bose-violating two-photon state will be $|B\rangle=N(|S\rangle+\mu |A\rangle)$ ($N$ is the normalization factor). Next, due to superposition principle, we can add to it the state $-N|S\rangle$, obtaining $|B'\rangle=N\mu |A\rangle$. Now, $|B'\rangle$ is obviously not properly normalized, and after normalization, it becomes just $|B''\rangle=|A\rangle$, i.e. ``amplitude for two photons to be in an antisymmetric state'' becomes 1 instead of $\mu$.

So  it is hard to give physical meaning to the concept of ``probability for two photons to be in an antisymmetric state'', unless superposition principle is modified in some way.

2. It is believed \cite{dbd} that the `Bose' and `Fermi' amplitudes $A_B$ and $A_F$ do not interfere, i.e., the total probability is assumed to be $\Gamma\propto |A_B|^2+\nu |A_F|^2$ rather than $\Gamma\propto |A_B+\mu A_F|^2$. In Ref. \cite{dbd} this assumption is justified by invoking the rule that the matrix elements of a symmetric Hamiltonian has zero matrix elements between the states of different symmetry \cite{ap}. However, this rule appears to be superfluous as the model \cite{dbd} itself dictates whether $A_B$ and $A_F$ interfere or not. Indeed, in this model, the necessary and sufficient condition of non-interference is $|A_{12}|=|A_{21}|$ (assuming for simplicity that $\mu$ is real).
(It should be noted that whether the superposition is coherent or incoherent is irrelevant for the decay $Z' \rightarrow \gamma \gamma$, which is forbidden at zeroth order.)

3. Should we consider $\nu$ is a `universal' parameter, i.e., independent of the physical process, energy scale etc.? This may or may not be true depending on the details of the underlying specific theory. For instance, it is not inconceivable that $\nu$ could be energy-dependent and grow with energy. In this case, upper limits on $\nu$ obtained in low- and high-energy processes would not be directly comparable to each other.

4. Finally, Ref.\cite{dbd} makes (implicitly) a  strong but arbitrary conjecture that the amplitude $A_{12}$ should be calculated {\em in the Standard Model}. 
There is nothing wrong with it as long as we remember that:\\
---this is just one of many possible assumptions, and certainly not the most general or unique.\\
---the parameter $\nu$ is meaningful only if this assumption is made.

5.  If we pursued a similar approach, the result would be that  the rate of $Z' \rightarrow \gamma \gamma$ decay becomes rather  small: due to a fermion loop it would be suppressed by at least a factor of $\alpha^2$ (cf. \cite{dbd}). As a result, its search at LHC  would require a higher sensitivity.

 By contrast, Ref. \cite{ign}  did not rely on this or similar conjectures, but tried to be more general. In this general approach, there is no point in estimating $\nu$, because this parameter belongs to a different model based on additional specific assumptions. 
 
 We feel that at the moment our level of understanding is, unfortunately, 
insufficient for telling on purely theoretical grounds which approach is correct, and we need an experimental search that could settle the issue.
\section{The  $Z'$  decays }
Consider now the new  boson $Z'$, which 
 appears in many models of physics beyond the SM, see e.g. \cite{review}-\cite{kozlov}. The $Z'$ is assumed to be a 
 more massive than the gauge boson $Z$ of the standard model. 
The most direct channel to probe the existence of the $Z'$ at a hadron collider,
 such as the CERN LHC \cite{lhc},  is  the Drell-Yan process. The $Z'$  that decay to leptons,
 $pp \to Z' \to l^+ l^- +X$ with  $l= e, \mu$,   
have a simple, clean experimental signature, and potentially could be discovered at the 
LHC with a mass up to 5 TeV, see e.g.\cite{cms, atlas}. 
 This new object is supposed 
 to be neutral, colorless and self-adjoint, i.e., it is its own antiparticle. 
 The mass  of the new boson could  be identified  unambiguously by a study of a  resonance peak in the dilepton 
invariant mass distribution.
The $Z'$ may be classified according to its spin, which 
 could be defined  by measuring the 
dilepton angular distribution in the reconstructed $Z'$ rest frame.
The $Z'$  could be  a spin-0 
$\overline{\nu}$ in R-parity violating SUSY, a spin-2 Kaluza-Klein (KK) excitation of the graviton as in the 
Randall-Sundrum (RS) model, or even a spin-1 KK excitation of a SM 
gauge boson from some extra dimensional model \cite{tr}. Another possibility for 
the spin-1 case is that the $Z'$ is a true $Z'$, i.e. a new 
neutral gauge boson, which is  the carrier of a new force, 
arising from an extension of the SM gauge group.  For 
much more extensive discussions of specific models and other implications see  several excellent reviews \cite{pl, tr, al, mc, dittmar},
and  a more complete list of references therein. 

The current best direct experimental lower limits on the  mass of $Z'$ of a few popular models came from the Tevatron and  restrict the $Z'$ mass to be greater than about 900 GeV when its couplings to SM fermions are identical to  those of the Z boson \cite{pdg}. 
  
Consider now the allowed branching fraction  $Br(\zptg)=\Gamma(\zptg)/\Gamma(Z'\to l^+l^-)$    in several interesting
 $Z'$ models.  First, we shortly 
describe these models and the SM fermions couplings the $Z'$. 

\begin{itemize}

\item the E$_6$ models   are described by the breaking chain
$E_6 \to SO(10)\times U(1)_\psi \to SU(5)\times U(1)_\chi \times U(1)_\psi
\to SM\times U(1)_\beta $. Many studies of $Z'$ are focusing on the two 
extra $U(1)'$ which occur in the above decomposition of the $E_6$.
The lightest $Z'$ is defined as :
\begin{equation}
Z'= Z'_\chi cos\beta  + Z'_\psi sin\beta 
\label{zpr}
\end{equation} 
where the values $\beta = 0$ and $\beta= \pi/2$ corresponds to 
pure $Z'_\chi$ and $Z'_\psi$ states of the $\chi$- and $\psi$-model,
 respectively. 
The value $\beta= arctan(-\sqrt{5/3})$
is related to a $Z'_\eta$ boson that would originate from the direct breaking 
of $E_6$ to a rank-5 group in superstrings inspired models.

\item  the  Left-Right
Symmetric (LRSM) model  is based on the symmetry group 
$SU_C(3) \otimes SU_L(2) \otimes SU_R(2)\otimes U(1)_{B-L}$ \cite{lr}, 
in which $B$ and $L$ are the baryon and lepton numbers, respectively.
The model necessarily incorporates  
three additional gauge bosons $W^\pm_R$ and $Z'$.
The most general $Z'$ is coupled  to a linear combination of right-handed 
and $B-L$ currents:
\begin{equation}
J^\mu_{LR}=\alpha_{LR}J^\mu_{3R}-(1/2\alpha_{LR})J^\mu_{B-L}
\end{equation}
where $\alpha_{LR}=\sqrt{(c_W^2g_R^2)/s_W^2g_L^2)-1}$, with 
$g_L=e/s_W$ and $g_R$ are the $SU(2)_L$ and $SU(2)_R$ coupling
constant with $s_W^2=1-c_W^2=sin^2 \Theta_W$. The $\alpha_{LR}$-parameter 
is restricted to be in the range $\sqrt{2/3} \lesssim \alpha_{LR} \lesssim \sqrt{2}$. The upper bound corresponds to the so-called manifest LRSM with $g_l=g_R$,
 while 
the lower bound corresponds to the $\chi$-model discussed above, since 
$SO(10)$ can results to both $SU(5) \times U(1)$ and $SU(2)_R \otimes SU(2)_L \otimes U(1)$ breaking parameter. 
To simplify our study, we will use further the following standard assumptions:
(i) the mixing angles are small; 
(ii) right-handed CKM matrix is identical to the left-handed one, and  
(iii) $g_R=g_L$.

\item in the sequential model (SSM) the corresponding $Z'$ boson has the 
same couplings to fermions as the $Z$ of the SM. The $Z'$ could be 
considered as an excited state of the ordinary $Z$ in models with 
extra dimensions at the weak scale.    

\item the Stueckelberg extension of the SM (StSM) is based on the gauge group
$SU(3)_C \otimes SU(2)_L \otimes U(1)_Y\otimes U(1)_X$ \cite{stueck}.
This extension of the SM involves a mixing 
of the $U(1)_Y$ hypercharge gauge field and the $U(1)_X$ Stueckelberg gauge field. The 
Stueckelberg gauge field has no couplings to the visible sector fields, while it may 
couple to a hidden sector, and thus the new physical $Z'$ gauge boson connects with the 
visible sector only via mixing with the gauge bosons of the physical sector. These mixings, 
however, must be small because of the LEP electroweak constraints and consequently the 
couplings of the $Z'$ boson to the visible matter fields are extra weak, leading to a very 
narrow $Z'$  resonance. The width of such a boson could be as low as a few MeV or even 
lower. An exploration of the Stueckelberg $Z'$  boson in the 
Tevatron data was recently carried out in \cite{stueck1}. 
 Such a resonance may also be  detectable via the Drell-Yan process at the LHC by an analysis of a dilepton pair arising from the $Z'$ decay \cite{stueck2}. 
The coupling structure of the Stueckelberg $Z'$ gauge boson with 
visible matter fields is suppressed by small mass mixing parameters thus leading to a very 
narrow $Z'$ resonance. Below we will assume that the $Z'-Z$ mixing strength is $\epsilon = 0.06$ \cite{stueck1}. 
\end{itemize}

The $Z'$ boson partial decay width into a fermion-antifermion pair 
 is given by 
\begin{eqnarray}
&&\Gamma(Z' \to f\overline{f})=N_C\frac{\alpha M_{Z'}}{6c_W^2}\sqrt{1-4\eta_f}\times \nonumber\\
&&[ (1+2\eta_f)(g^{f}_L)^2+(1-4\eta_f)(g^{f}_R)^2 ]
\end{eqnarray}
where $N_C$ is a color factor ($N_C=3$ for quarks and $N_C=1$ for leptons),
$g^{f}_L, g^{f}_R$ are the left- and right-handed couplings of the $Z'$ to 
the SM fermions,  $\alpha$ is the electromagnetic 
coupling constant, which is $\alpha \simeq 1/128$ at the $M_{Z'}$  scale, 
 and  $\sqrt{\eta_f}(=m_f/m_{Z`})$ is assumed to be  $\ll 1$.
The left-handed couplings of the $Z'$ to the SM neutrinos are 
$g^{f}_L = \frac{3cos \beta}{2\sqrt{6}}+\frac{\sqrt{10}sin\beta}{12}$ 
and $g^{f}_L = \frac{1}{2\alpha_{LR}}$ for E$_6$ and LRSM models, respectively,
while the right-handed couplings $g^{f}_R=0$ in both models.  
The $(g^{f}_L)^2$ is restricted to lie 
 in the range   
$0.07 (\beta\simeq \pi/2) \lesssim (g^{f}_L)^2 \lesssim 0.45~(\beta \simeq 0.4)$ for the
E$_6$ model \cite{e6}, and in the range  $ 1/8 \lesssim (g^{f}_L)^2 \lesssim 3/8$
for the LRSM model. The detail discussions of the StSM $Z'$ decay modes and couplings can be 
found in \cite{stueck2}.

 For the decay rate $\zptg$ in the $Z'$ mass range $M_{Z'}\lesssim 5$ TeV 
 one may expect 
\begin{equation}
\Gamma (Z' \rightarrow \gamma \gamma) \lesssim 1.4\cdot M_{Z'}[\rm TeV]~MeV, 
\label{zpw}
\end{equation}
as it follows from Eqs.(\ref{rate1},\ref{gb1}).

Due to the possible strong dependence of the couplings $g_B$ on the boson mass and the discussion at the end of Section II, the inequality (\ref{zpw}) should be viewed as an indication of the ballpark of the possible values of the width, rather than the firm limit.
The same applies to Fig.3 and caption to Table 1.
In Table 1 expected properties of $Z'$ bosons for  several models 
are summarized. The $\zptg$ decay rate is  calculated  by  taking into account the LEP limit on  
the Bose violation  coupling \eqref{gb1}. 

\begin{table}[h]
\begin{center}
\begin{tabular}{|c|c|c|c|}
\hline
model  & $\Gamma_{tot}/M_{Z'}$&$B(Z'\to \mu^+\mu^-)$ &$\frac{\Gamma(Z'\to \tg)}{\Gamma(Z'\to \mu^+\mu^-)}$ \\
        &  $\times 10^2$ & $\times 10^2$ & $\times 10^3$\\
\hline
\hline 
 $Z_{SSM}$ & 3.1 &3.2 & $\leq 1.5$ \\
\hline
 $Z_{\psi}$ & 0.6 &4.1 & $\leq 5.8$ \\
\hline
 $Z_{\eta}$ & 0.7 &3.5 & $\leq 5.8$ \\
\hline
 $Z_{\chi}$ & 1.4 &5.6 & $\leq 1.7$ \\
\hline
 $Z_{LRSM}$ & 2.2 &2.2 & $\leq 2.8 $ \\
\hline
 $Z_{StSM}$ & 0.006 & 12.3 & $\leq 23$ \\
\hline
\end{tabular}
\end{center}
\caption{Summary of expected properties of $Z'$ bosons for several models. The first column shows the 
ratio of the $Z'$ width to its mass $M_{Z'}$, the second column shows the dimuon branching fraction, and the 
third column gives the upper limit for the ratio of the $Z'$ diphoton and dimuon decay rates calculated
taking into account the limit of \eqref{gb1}. For interpretation of results, see comments after Eq. (22). }
\label{brrat}
\end{table}

\section{Signal and backgrounds}
Next,  let us explain how to search for the $Z'\to \tg $
decay at the LHC.  The $Z'$ decay into two photons is a rare decay mode, with a branching fraction
 $B(Z'\to \tg)$ in the range  of $\lesssim 10^{-4}-10^{-3}$. 
 The final state consists of two high $p_T$ photons, $p_T \simeq M_{Z'}/2$ ,  detected in a 
 LHC electromagnetic  calorimeter surrounding the $pp$ collision region
\cite{cms, atlas}.  The experimental signature of  the $Z'\rightarrow \tg$ decay  
is a peak in the invariant mass distribution of these  photons over the continuum background.
An important point is that if the  decays of the $Z'$  into leptons occur, 
the position of the $Z'$ mass peak and its expected experimental width
in the diphoton invariant mass distribution can be predicted by the analysis of on-peak data based on the observation of a leptonic   $Z'$ decay mode. 
The allowed maximal branching fraction $\Gamma (\zptg)/\Gamma(\zptmu)$ 
 calculated for the Bose violating coupling constant  $g_{B1}=1.0\times 10^{-2}$ taking into account 
\eqref{zpw} is shown in Table \ref{brrat}. One can see, that, for example, for  the  $g_B\simeq 10^{-2}$, the process $\zptg$ could amount to more then 1\% of the total muonic $Z'$ decay rate in the StSM model. Hence, 
 if $Z'$ is observed at the LHC the search for its diphoton  decay mode  is  of
 great interest for possible  observation of Bose symmetry violation. 
\begin{figure}[tbh!]
\begin{center}
    \resizebox{8cm}{!}{\includegraphics{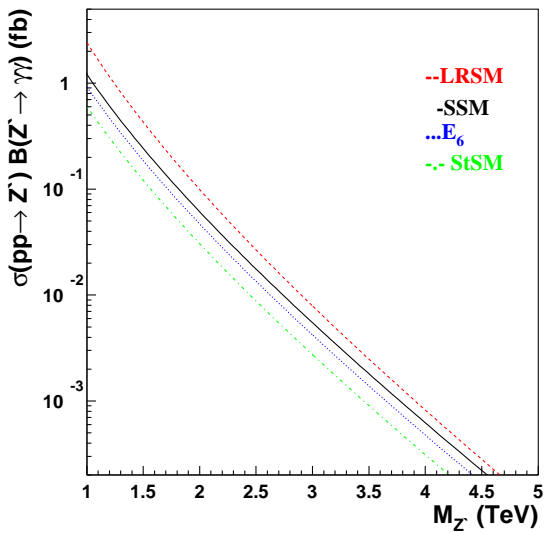}}
     \caption{The production cross section $\sigma(pp\to Z')\cdot B(\zptg)$ at the LHC at $\sqrt{s}=14$ TeV in several $Z'$   
        models indicated in the plot , 
     calculated as a function of the $Z'$ mass assuming the value of the Bose violating coupling constant $g_{B1}=1.0\times 10^{-2}$.}
\label{crsec}
\end{center}
\end{figure} 

For the appropriate $Z'$ coupling constants discussed above, 
the  production cross section $\sigma(pp\to Z')\cdot B(\zptg)$ at the LHC for several $Z'$ models are calculated  
in the framework of PYTHIA \cite{pith}. In this evaluation, the default CTEQ5L parton distribution functions \cite{pdf} are used with no K-factors included. 
 In most of our analysis we also
neglect errors associated with imprecise knowledge of parton distribution functions;  the related systematic errors, however,
 will be included into the final results: see Sec. VI.
 Fig.\ref{crsec} shows the production cross section $\sigma(pp\to Z')\cdot B(\zptg)$ at the LHC at $\sqrt{s}=14$ TeV 
 in the E$_6$, LRSM and St\"uckelberg models  
     calculated as a function of the $Z'$ mass assuming the value of the coupling constant $g_{B1}=1.0\times 10^{-2}$. The StSM curve is calculated for  the $Z'-Z$ mixing strength value $\epsilon = 0.06$.
 Interestingly, although the StSM production cross section $\sigma(pp\to Z')\cdot B(Z'\to l^+ l^-)$ is an order 
of magnitude below those of other $Z'$ models, the cross section $\sigma(pp\to Z')\cdot B(\zptg)$ is comparable
to the corresponding cross sections in other models.   

Although the  experimental  signature of the decay $\zptg$ at the LHC is expected to be relatively 
clean, in order to discover  this process  
 one has to determine  whether the $Z'$ mass peak in the diphoton invariant mass spectrum
  could be distinguished from the background due to the standard model reactions.     At a hadron collider experiment
the diphoton production with a large invariant mass is 
a well known and studied  background not only for the search of the two photon  decay of the 
Higgs boson, but  also for searches of new  
heavy resonances, extra spatial dimensions, or cascade decays of 
heavy new particles \cite{cms, atlas} where it is a source of significant background. 
The dominant standard model background sources  to our signal are (see e.g.\cite{cms}):
 
\begin{itemize}

\item the prompt $\tg$ production either form the quark annihilation or gluon fusion. As the final states from our signal 
and from these processes are identical, this   is irreducible intrinsic background. 

\item  The $\gamma +$jets production consisting of two parts: 
(i) prompt photon from hard interaction plus the second 
photon coming from the outgoing quark due to initial and final state radiation and 
(ii) prompt photon from hard interaction plus the decay of a neutral hadron (mostly isolated) 
in a jet, which could fake a decay photon. 
The $\gamma + jet$, $pp\to \gamma+ jet+X$ with a jet faking photon production turned out to be one of the most important backgrounds.

\item The background from QCD hadronic jets, consisting of quarks that fragment into a 
high momentum $\pi^0$ , which subsequently decays as $\pi^0 \to \tg$. 
The resulting photon showers may overlap, and can 
pass the photon selection.

\item  other possible sources of background is the Drell Yan  productions 
$pp \to e^+ e^- +X$.
The production of a high-transverse-momentum lepton pair, can lead to a diphoton final state 
if both electrons produce hard bremsstrahlung photons, or if the electron tracks fail to be 
properly reconstructed. We, however, assume a high efficiency of a LHC detector silicon tracker veto
which, being applied to  photon-containing  candidate events,  makes this background negligible. 
\end{itemize}

\section{Simulations of the process  $pp\to \zptg$ at $\sqrt{s}=14$ TeV}

To make quantitative estimates, 
we performed simplified simulations at the generator level of the $Z'$ production 
 followed by the decay $\zptg$ in the reaction $pp \to Z'$ 
 and the corresponding background processes at the LHC. We  consider, as an example,  the CMS detector  \cite{cms}. 
As  the signal events preferentially
populate the large transverse momentum part of 
the phase space, events were generated with $P_T> 100$ GeV 
(CKIN(3) parameter) and $|\eta| <2.7$  respectively. This allows us to reduce 
the time of computations and 
also to exclude of a very large fraction of the standard model 
events, which are peaked at  small transverse momenta. 

The CMS  detector is described in detail in Ref. \cite{cms}. 
It consists of several subsystems: a superconducting magnet, a Si-tracker
surrounded by an electromagnetic calorimeter followed by a hadronic calorimeter and muon chambers 
used for the detection and reconstruction of the events.
The CMS experiment  
uses lead tungstate  
crystals for the electromagnetic calorimeter (ECAL). 
Each crystal measures about 22 $\times$ 22 mm$^2$ 
and covers $0.0175\times 0.0175$ (about $1^o$ ) in the $\Delta \eta - \Delta \phi$
space ($\phi$ being the azimuth angle).

For photon reconstruction at the generator level, we have used the ``hybrid" clustering algorithm,
to account for also fake photons arising from jets \cite{cmstdr}.  
Photon candidates are reconstructed as superclusters in the CMS electromagnetic calorimeter 
(ECAL), within the fiducial regions of the barrel (EB) 
$|\eta| < 1.4442$ and endcaps (EE) $1.566 < |\eta|  < 2.5$. 
The superclusters are extended in 
$\phi$ to recover the energy deposited by electron bremsstrahlung and photon conversions.
We consider a photon in the ECAL as a local deposition of 
electromagnetic energy by electrons or photons 
 contained in a cone  
$R= \sqrt{\Delta \eta^2 + \Delta \phi^2}< 0.09$ with no associated tracks.
 This definition 
is equivalent to $10\times 10$ crystal size in the CMS detector. 
The CMS experiment uses $5\times 5$ crystal size to form an 
energy cluster to reconstruct a photon candidate. However, in 
our efforts to mimic this reconstruction process at the generator level, 
we choose to be conservative and use only a $10\times 10$ crystal.
The momentum of the photon candidate is defined as the vector sum of 
 the momenta of the electromagnetic objects in such a crystal. 

As mentioned above, the main challenge to identifying the true photon candidates arises from jets
 faking photons, see e.g.  \cite{cms}. This 
 occurs when a jet from the standard model processes with  $\gamma + jet$ or  $jet-jet$ final state 
 is dominated by a neutral hadron, such as, for example,  a $\pi^0$ or $\eta$,
  which decays into two 
 photons. If the hadron is highly energetic, so that the cosine of the opening angle between  
the  two decay photons is $cos(\Theta_{\gamma_1 \gamma_2})> 0.9$, this angle is  difficult to 
resolve  and the photons can be misidentified as a single energetic photon. 
  To suppress  such 
backgrounds, we use  various isolation variables,  without, however,  taking into account 
such photon object characteristics as the lateral and longitudinal electromagnetic shower shape.
Jets typically have a larger number of  
charged particles reconstructed in their vicinty, and also a larger ratio of hadronic to electromagnetic deposited
 energy  than photons. Likewise, hadronic and electromagnetic deposits  
arising from jets will be less isolated than for photons.
Fake photon signals arising from a jet can be rejected by requiring 
either the absence of charged tracks above a certain minimum transverse momentum($P^{tr}_{T min}$ ) 
associated with the photon or the absence of additional energetic particles 
in an annular cone ($R_{iso}$) around the photon candidate. Following the  diphoton analysis similar to Ref.\cite{ind},
we  have considered two variables for the isolation purposes: (i) the number of tracks ($N_{tr}$) from
charged particles, such as $\pi, K, p, ...$,
inside a cone around the photon and (ii) the scalar sum of transverse energy 
($E_{T}$) inside a cone around the photon. To identify the photons from the decay $\zptg$,
 the following  $\gamma$ events selection criteria  are used: 
\begin{itemize}

\item $P^{\gamma 1}_T \geq 100$ GeV, $P^{\gamma 2}_T \geq 100$ GeV;

\item $|\eta^{\gamma 1 ,\gamma 2}| < 2.5$, $|\eta^{\gamma 1 ,\gamma 2}| \neq 1.4442-1.5666$;

\item $cos(\Theta_{\gamma 1 ,\gamma 2})\leq 0.9$;

\item $N_{tr}= 0$ for $P^{tr}_T \geq 3.0$ GeV within $R_{iso}\leq 0.35$;

\item $E_{T}< 5.0$ GeV within $R_{iso}\leq 0.35$;

\end{itemize}
Using this algorithm 
and requiring the photon to be isolated,  the estimated probability of a jet faking a photon 
in $\gamma +j et$ channel is $\simeq 10^{-4} - 10^{-3}$. The major sources of 
fake photons are $\pi^0$ ($\gtrsim 80\%$), 
with  the  rest coming from other  sources.  

As the next step, the signal event candidates are selected by requiring that the final state contain
at least 
 two or more isolated photons and a jet(s). Events are studied in which either both photons are in 
the barrel  calorimeter, or one photon is in the barrel  and the other is in the 
end cup  calorimeter.  
The diphoton invariant mass distribution  is calculated 
for two highest $P_T$ photons and 
  histogrammed in bins equivalent to the mass resolution. 
  The combined acceptance and selection efficiency for  events with $P_T > 100$ GeV and 
 for  $M_\zp > 1000$ GeV is found to be $\gtrsim 70\%$.

\section{Results}
\begin{figure}[hbtp]
\begin{center}
\includegraphics[width=0.5\textwidth]{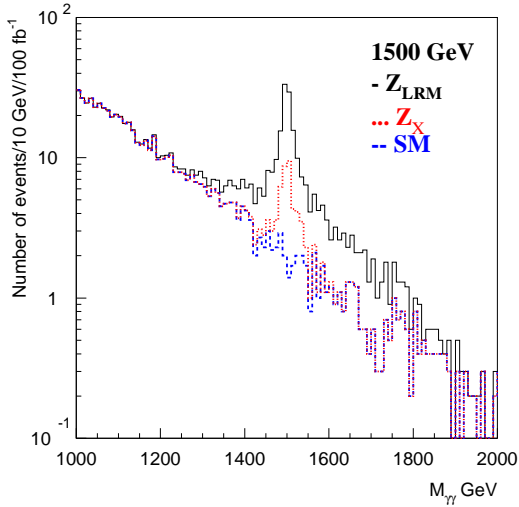}
     \caption{The diphoton invariant mass $M_{\gamma \gamma}$ distribution in the presence
of the SM background calculated for the LRSM and E$_6(Z_\chi)$ models
for the $Z'$ with the mass of 1.5 TeV and 
the LHC integrated luminosity $L=150~fb^{-1}$. 
 The Bose violating 
coupling is assumed to be $g_{B1}=1.0\times 10^{-2}$. }
      \label{plot}
   \end{center}
\end{figure}

 In Fig.~\ref{plot}  the invariant mass $M_{\tg}$ distributions   in the presence
of the standard model background are shown  for events simulated for the LRSM and  E$_6$($Z_\chi$) models
for the $Z'$ with the mass of 1.5 TeV, the Bose violating 
coupling  $g_{B1}=1.0\times 10^{-2}$, and 
the LHC integrated luminosity $L=150~fb^{-1}$. 
The additional broad peak and long tails below the $Z'$ peak are from  a combinatorial
background due to the wrong choice of photons. 
For the invariant $\tg$-mass $\gtrsim$ 1 TeV, the background under the $Z'$ peak drops quickly. 

The significance of the discovery  of the $\zptg$ events can be estimated as  \cite{nk}:
\begin{equation}
S=2(\sqrt{N_S+N_B}-\sqrt{N_B}),
\end{equation}
where $N_S$ and $N_B$ are the numbers of signal and background events
respectively, which pass the selection criteria described above.
These numbers of events are estimated from a search for the $\zptg$ mass peak, 
which was performed in the following way. For every $Z'$ mass value, 
the region around it in the $M_{\gamma \gamma}$ distribution  was fitted 
with a parametrized signal shape centered at the $M_{\gamma \gamma}$ value and superimposed 
over a polynomial background. 
 The normalization of 
each component is allowed to float in the fit. This procedure
is also used for the background estimate as a function of $M_{\gamma \gamma}$, 
with statistical uncertainties propagated from the fits. The systematic errors 
discussed below are also propagated to the fit procedure. 
The discovery potential of the $\zptg$ decay with the CMS detector is estimated 
assuming $S\gtrsim 3$. 

\begin{figure}[tbh!]
\begin{center}
    \resizebox{8cm}{!}{\includegraphics{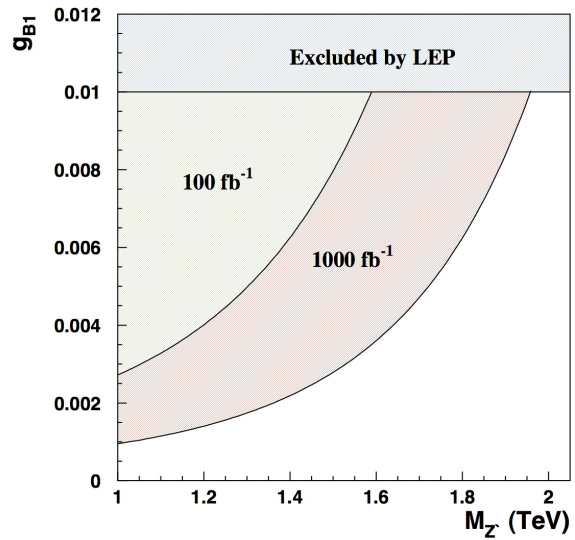}}
     \caption{The estimated CMS  discovery potential  in the $(g_B, M_{Z'})$ 
     parameter space calculated for the LRS model 
  and for the integrated LHC luminosity 100 and 1000 fb$^{-1}$. For interpretation of results, see comments after Eq. (22).}
  \label{limits}
\end{center}
\end{figure}

The final results of this analysis are presented  in Fig.\ref{limits},
 where the most stingent limits  for the coupling constant $g_B1$  
obtained for the LRS model are shown  as a function of  $Z'$ mass.
For the total luminosity of 100 fb$^{-1}$ and the mass about 1 TeV, the limits are about one order of magnitude 
lower than the corresponding present limit from the search for $Z\to \gamma \gamma $ decay 
mode at LEP, thus making the process $Z'\to \tg$ feasible for 
observation at the LHC.
For the higher integrated luminosity of 1000 fb$^{-1}$, the use of 
$pp\to Z' $ reaction allows to probe the decay $\zptg$ for
the $Z'$ masses up to 2 TeV.

 The systematic errors in the number of expected $\zptg$ events coming from various background uncertainties
are small since the background itself is rather small and the discovery
region is usually limited by the fast drop of the signal cross section
at high $Z'$ mass. 
The largest systematic uncertainties of the expected number of $Z'$-bosons arise from 
the luminosity measurement (6\%) and the choice of PDF 
(5\%). The latter uncertainty is determined from the variation in the efficiency 
when employing different PDF parameterizations. 
To study the effect of the detector energy
resolution on this analysis, the energy of the photons was smeared with the 
stochastic term of the CMS electromagnetic calorimeter 
energy resolution \cite{cmstdr}.
Due to limitations in computing time, we did not fully simulate the background 
from jet-jet events. Although the dijet cross section is quite large, 
given the low probability of 
a jet faking a photon  it is found  that the kinematical and isolation 
cuts used above reject the dijet background substantially \cite{ind}.
To get conservative estimate, we include  uncertainties in the background
estimate of 15\%. Finally, the error of the measured LHC integrated 
luminosity is taken to be 3\% \cite{cmstdr}.

\section {Conclusion}

To summarize, 
we consider a phenomenological  model of the Bose statistics violation.
We show that if a new heavy $Z'$ boson is observed at LHC, further  
searches for  the $Z'\to \tg$ decay mode would suggest an 
interesting additional direction to probe Bose symmetry violation at the high energy frontier.
We  have  demonstrated  that the discovery  regions in the 
($ g_{B1}; M_{\tg}$) parameter space for the  $\zptg$ decay substantially extend the 
excluded region from the CERN LEP. 
 The low-energy experiment in atomic spectroscopy  might be a sensitive  probe 
 of Bose symmetry violation  that is complementary to  collider experiments. 
 
\begin{acknowledgments}
We thank D.S. Gorbunov and  N.V. Krasnikov for helpful discussions, and M.M. Kirsanov 
for help in simulations and comments on the CMS ECAL photon isolation algorithm.
\end{acknowledgments}


\begin{thebibliography}{99}
\bibitem{e6}
The minimal possible  value for  $(g^{f}_L)^2$
in this model is zero , we take it to be $\simeq 0.07$ for $\beta=\pi/2$
 to have non-zero $Z'-\nu$ coupling.
\end{thebibliography}
\end{document}